\documentclass{moriond}

\def\be{\begin{equation}}
\def\ee{\end{equation}}
\def\bea{\begin{eqnarray}}
\def\eea{\end{eqnarray}}

\newcommand{\Photo}{\includegraphics[height=35mm]{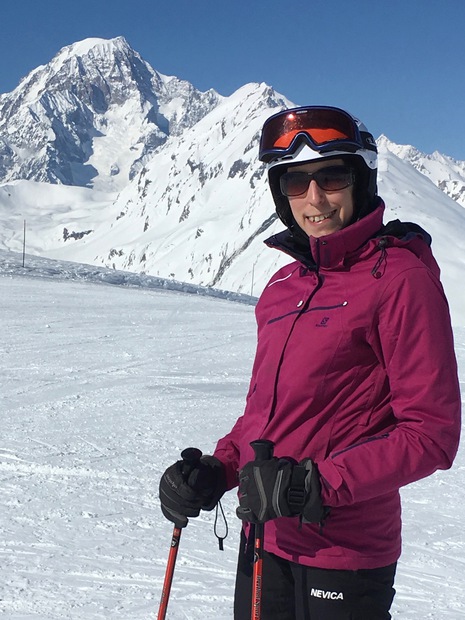}}

\begin{document}
\vspace*{4cm}
\title{THE C-BAND ALL-SKY SURVEY (C-BASS)}

\author{ ANGELA C. TAYLOR for the C-BASS Collaboration \\}

\address{Sub-department of Astrophysics, University of Oxford, \\
Denys Wilkinson Building, Keble Road, Oxford, OX1 3RH, UK}

\maketitle\abstracts{The C-Band All-Sky survey (C-BASS) is an experiment to image the whole sky in intensity and polarization at 5~GHz. The primary aim of C-BASS is to provide low-frequency all-sky maps of the Galactic emission which will enable accurate component separation analysis of both existing and future CMB intensity and polarization imaging surveys. Here we present an overview of the experiment and an update on the current status of observations. We present simulation results showing the expected improvement in the recovery of CMB and foreground signals when including  C-BASS data as an additional low-frequency channel, both for intensity and polarization.  We also present preliminary results from the northern part of the sky survey. }

\section{Introduction}

Detecting the primordial $B$-mode signal in the polarization of the cosmic microwave background (CMB) is one of the key goals in observational cosmology. However, the $B$-mode signal is so small that it is likely to be fainter than Galactic foregrounds even in the quietest patches of sky. Foreground subtraction will therefore be the limiting factor in the search for $B$-modes rather than sensitivity. To distinguish the CMB from Galactic foregrounds, wide frequency coverage is needed either side of the foregrounds minimum at 70 -- 100 GHz. Observations at the lower end of this range are more easily done using radiometer techniques than using bolometers. The C-Band All-Sky Survey (C-BASS) \cite{ProjectPaper} is a project to survey the whole sky at 5 GHz, at the lower end of the foregrounds frequency range, where synchrotron radiation is the dominant emission mechanism. This frequency is low enough that polarized synchrotron emission is detectable across almost all the sky, but high enough that Faraday rotation is a small effect apart from very close to the Galactic plane. By providing much better measurements of the synchrotron component, C-BASS data will help break degeneracies between fits to the CMB and all the foreground components, and thus improve our measurements of the CMB.     

\section{The instrument and survey}

\begin{figure}[t!]
\begin{center}
\includegraphics[height=3.5cm]{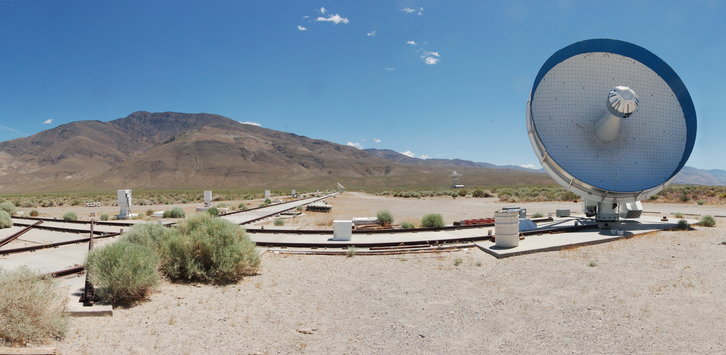}
\includegraphics[height=3.5cm]{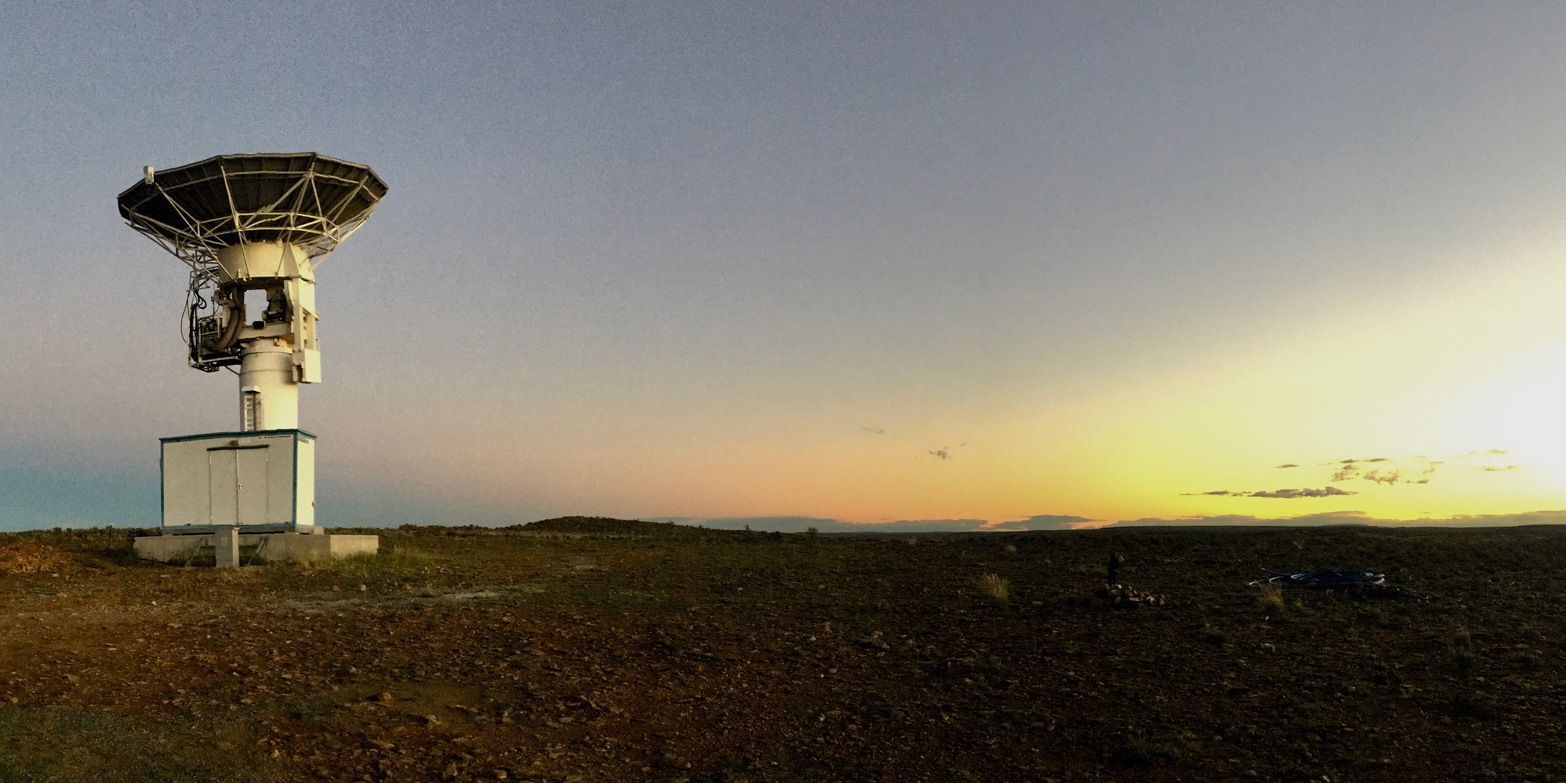}
\end{center}
\caption{(Left) The 6.1-m C-BASS North telescope at Owens Valley Radio Observatory. (Right) The 7.6-m C-BASS South telescope at Klerefontein, South Africa. Note that the horizon at C-BASS South is much flatter, which is expected to lead to less scan-synchronous groundspill. The overlap in sky-coverage between the two telescopes, and the different characteristics of their ground spillover signals, will allow us to partly break the degeneracy between ground emissions and sky modes that are symmetric about the celestial poles.}
\label{fig:telescopes}
\end{figure}

C-BASS consists of two telescopes (see Fig. \ref{fig:telescopes}), one in each hemisphere, with matched beam sizes and similar receivers providing intensity and polarization measurements in the frequency band 4.5 -- 5.5 GHz.\cite{ProjectPaper,Holler2011} C-BASS has a resolution of 45 arcmin, and maps the sky with no explicit filtering on angular scales larger than this. The survey sensitivity of better than $100 \, \mu$K-degree corresponds to a sensitivity to synchrotron emission at 100 GHz of $0.75\, \mu$K-arcmin (assuming a synchrotron spectral index $\beta \sim -3$), and is thus comparable to the most sensitive planned experiments at those frequencies. C-BASS will therefore provide low-frequency measurements sensitive enough for comparison with both current and future generations of CMB experiments.

C-BASS North has an analogue receiver \cite{King2014} which provides a single frequency channel covering the whole passband. C-BASS South has a digital backend which provides 128 frequency channels across the 4.5 -- 5.5 GHz band. As well as improving discrimination against radio frequency interference, this allows additional frequency coverage in the data analysis, in particular the measurement of Faraday rotation due to the Galactic magnetic field.  

\section{Simulations of the impact of C-BASS data}

\begin{figure}
\begin{center}
\includegraphics[width=0.49\textwidth]{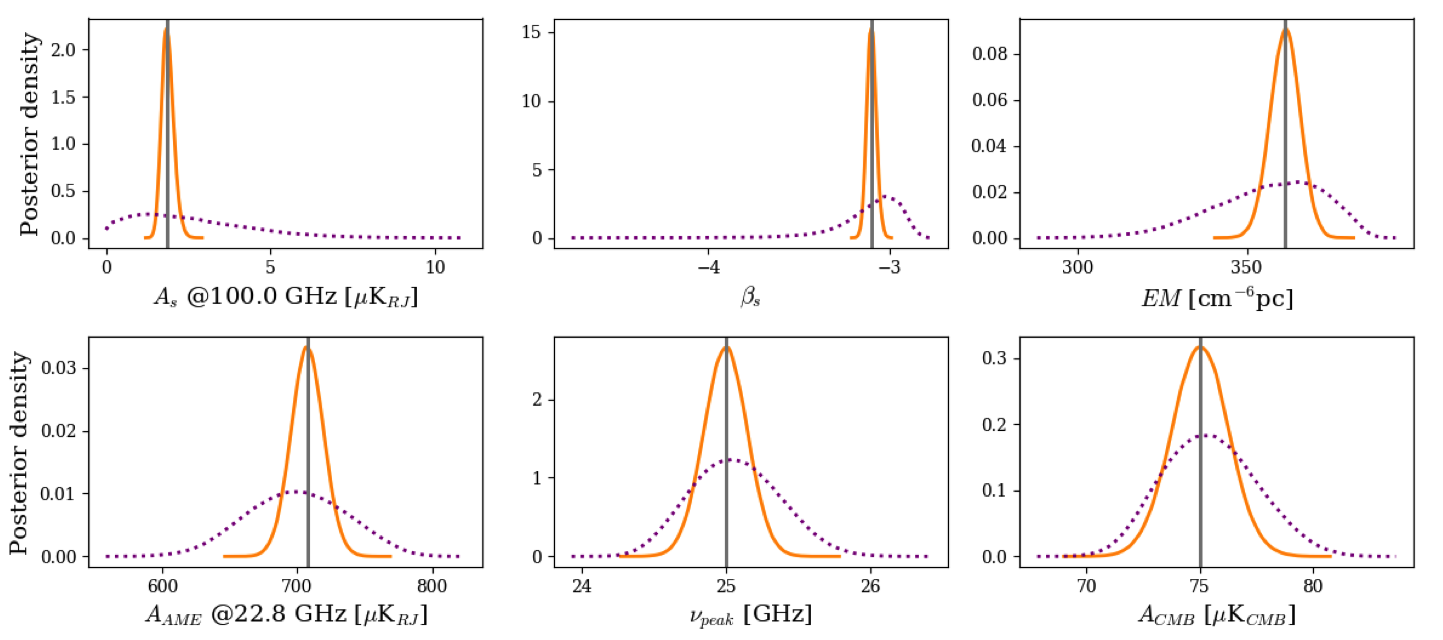}
\includegraphics[trim=0 -50 0 0 , clip,width=0.49\textwidth]{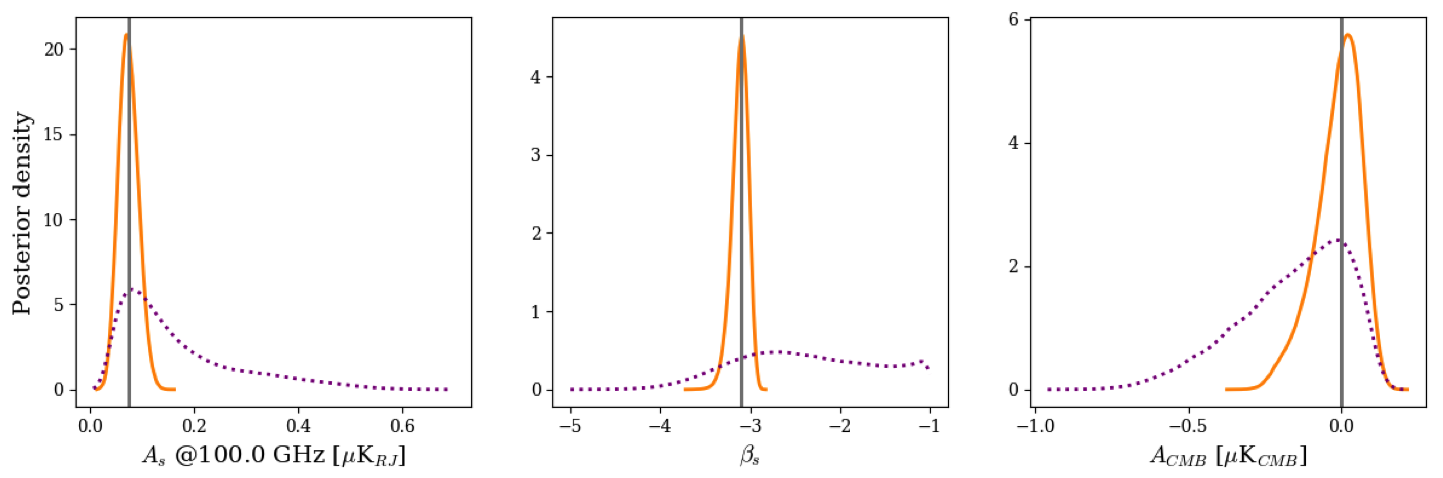}
\end{center}
\caption{\small Improvements expected in the recovery of CMB and foreground signals without (dotted purple) and with (solid orange) C-BASS data.  (Left) Posterior probability densities for recovered parameters (synchrotron amplitude and spectral index, free-free emission measure, AME amplitude and peak frequency, CMB amplitude) in intensity using current data sets (Planck and WMAP) in a high-foregrounds (low Galactic latitude) pixel. (Right) Posterior probability densities for synchrotron amplitudes and index, and CMB amplitude, in polarization in a low-foreground pixel using sensitivities from Planck plus the proposed LiteBIRD satellite mission. Thermal dust parameters are omitted for clarity. In both cases the additional C-BASS data breaks degeneracies in the separation of CMB and foreground signals and significantly improves the accuracy of the CMB measurement. Figure adapted from Jones et al.$^1$}
\label{fig:sims}
\end{figure}

To quantify the way in which C-BASS data will improve the component separation problem in both intensity and polarization, we have simulated the recovery of the CMB signal and the foreground parameters in individual representative pixels. In such a single pixel, we can do a full Bayesian MCMC recovery of the CMB and foreground parameters based on mock observations with current and future CMB experiments. Full details and further examples of this technique will be given in a forthcoming paper.\cite{LukePixelSims} Fig. \ref{fig:sims} shows some typical results in intensity using sensitivities from current data sets (Planck and WMAP), and in polarization from Planck and an early specification of LiteBIRD, \cite{LiteBIRDSpecs} both with and without C-BASS data. The sky is modelled with foregrounds consisting of a single power-law synchrotron component, free-free emission, anomalous microwave emission (AME) of variable peak frequency, and a modified blackbody thermal dust model, with only synchrotron and thermal dust assumed to be polarized. Even in this highly simplified model, the C-BASS data provide strong additional constraints on the synchrotron, free-free and AME contributions to the observed signals. This leads to significantly smaller uncertainty in the recovered CMB amplitudes. 

\section{Current status}

Observations with C-BASS North have been completed, while C-BASS South observations are ongoing. Below we show some preliminary results from a partial analysis of the C-BASS North data, using just the scans made at the elevation of the north celestial pole. Additional observations were also made at other elevations, which will result in significantly improved sensitivity, more uniform sky coverage, and better rejection of residual striping due to better cross-linking.   

\subsection{Intensity observations}

\begin{figure}[h]
\centering
  \includegraphics[trim=0 -20 0 0,clip,width=0.49\textwidth]{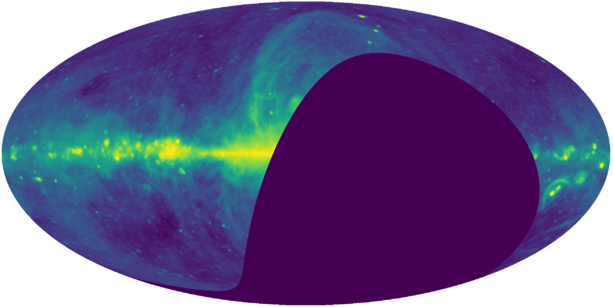}
  \includegraphics[width=0.49\textwidth]{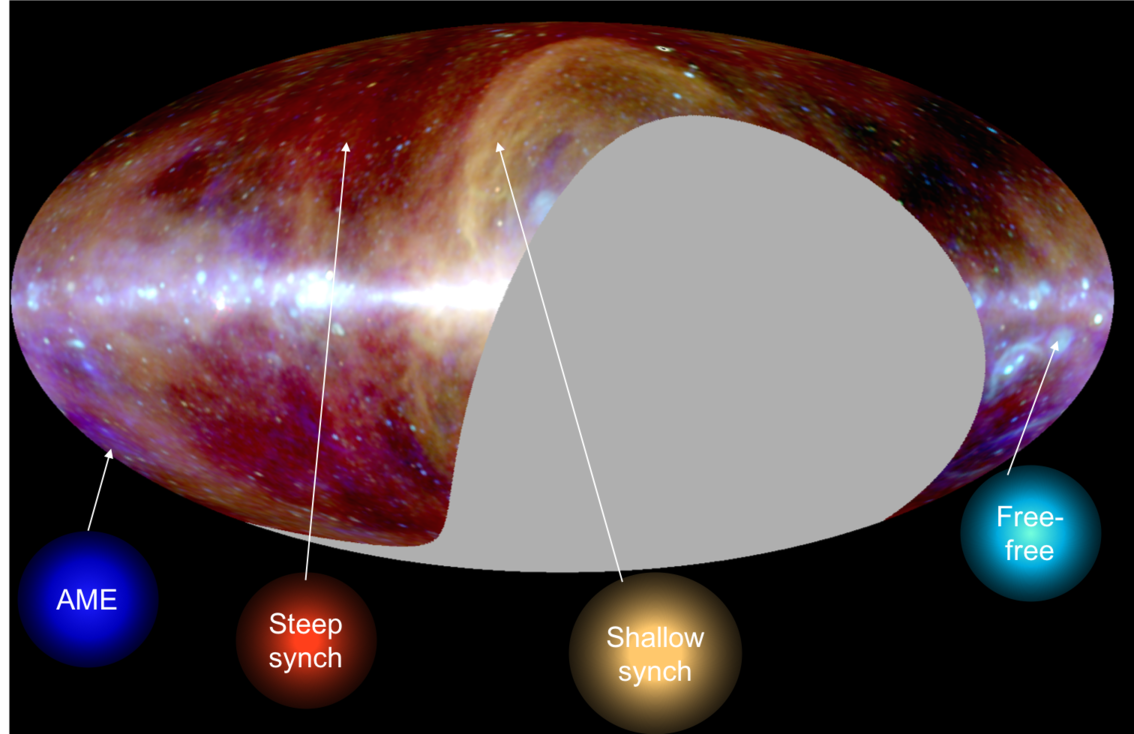}
  \caption{\small (Left) C-BASS North intensity map using just under half of the total data set. (Right) Three-colour image using the Haslam 408 MHz, C-BASS 5 GHz, and WMAP $K-V$ (effectively 23 GHz) maps for the red, green and blue channels respectively. The relative intensities of the three channels are chosen such that a temperature spectral index of $-2.7$ would appear as white. Adapted from Jew (2017).$^6$}
  \label{fig:IandThreeCol}
\end{figure}

Fig. \ref{fig:IandThreeCol} (left) shows the C-BASS North intensity image. To capture the dynamic range of the image, which ranges from a noise level of less than 0.1 mK to a peak of several K on the Galactic plane, an arctan intensity scaling is used. A variety of different null tests (e.g. day/night, seasons, alternate observations) show maximal residuals of a few mK at most.\cite{LukeThesis} A visual indication of the additional information available from combining C-BASS with existing surveys is shown in Fig. \ref{fig:IandThreeCol} (right), which combines the C-BASS intensity map with the Haslam et al. 408 MHz map~\cite{Haslam1982,Remazeilles2015} and the WMAP K-band map~\cite{WMAP} (corrected for the CMB contribution by subtracting the V-band map) in a three-colour image. The different spectral energy distributions of synchrotron radiation, free-free emission and AME result in these components appearing as distinct colours in the image.  

\subsection{Polarization observations}

\begin{figure}
\begin{center}
\includegraphics[width=0.46\textwidth]{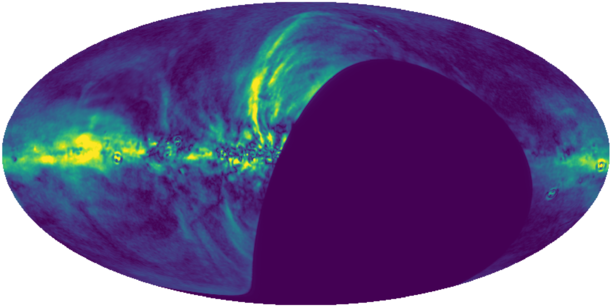}
\includegraphics[trim=0 40 0 0,clip,width=0.46\textwidth]{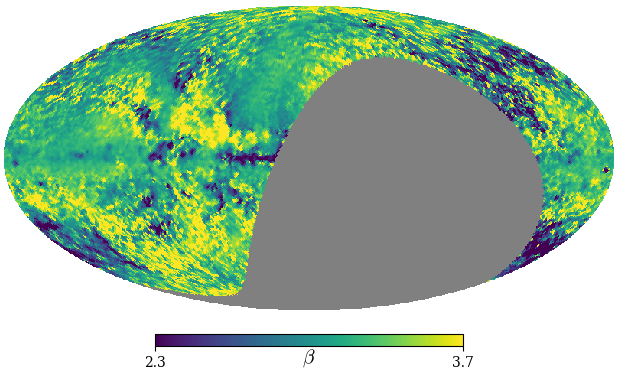}
\end{center}
\caption{\small Preliminary C-BASS North polarization results. (Left) The C-BASS polarized intensity map, on a linear colour scale from 0 to 40 mK. (Right) Spectral index map between C-BASS at 5 GHz and Planck at 30 GHz (linear colour scale from 2.3 to 3.7). Spectral index variations of up to $\pm 0.5$ dex are detected at high significance. Sensitivity is limited by the Planck noise level. Figures from Jew (2017).$^6$ \label{fig:pol-results}}
\end{figure}

The preliminary C-BASS North polarized intensity image is shown in Fig. \ref{fig:pol-results}.  Polarized intensity is detected with good signal-to-noise ratio over a large fraction of the sky. We can combine this map with the Planck 30~GHz polarized intensity map$\,$\cite{Planck2016a} to get an initial estimate of the polarized synchrotron spectral index over much of the sky.\cite{LukeThesis} This map was constructed from a pixel-by-pixel fit to the two input maps taking account of the non-Gaussian noise statistics of polarization amplitude, and is not biased by the noise power in either map. Some of the structure in the spectral index map is due to depolarization which is clearly visible in the C-BASS map close to the Galactic plane. However, much of the spectral index variation across the sky appears to be genuine.  Similar spectral index variations are also observed between S-PASS at 2.3 GHz and the lower WMAP and Planck frequencies.\cite{SPASS}  These variations are very significant for component separation at the level of accuracy required for deep $B$-mode searches. As well as showing that the synchrotron component cannot be fitted using a single global spectral index, small-scale variations such as this will result in the effective spectrum on large scales being curved,\cite{Chluba2017} effective modelling of which will require more frequencies to be observed. 

\section{Conclusions}

C-BASS is providing sensitive and accurate measurements of the sky in intensity and polarization at 5 GHz. Simulations confirm that these will be essential for CMB foreground component separation for deep $B$-mode searches. Preliminary data from the northern part of the survey also show that there are significant variations in synchrotron spectral index across the sky. Observations in the south are ongoing and will also provide frequency resolution within the C-BASS band.




\end{document}